\documentclass{sig-alternate-05-2015}

\begin{document}

\setcopyright{acmcopyright}

\doi{---}

\isbn{---} 

\conferenceinfo{MSWiM '16}{November 13--17, 2016, Malta}

\acmPrice{\$15.00}

%
\conferenceinfo{MSWiM '16}{November 13--17, 2016, Malta}

\title{Learning-Based Resource Allocation Scheme for TDD-Based CRAN System}
\subtitle{ }
%
%
%
%
%

\numberofauthors{5} 
%
\author{
%
%
\alignauthor
Sahar Imtiaz\\
       \affaddr{KTH Royal Institute of Technology,}\\
       \affaddr{School of Electrical Engineering,}\\
       \affaddr{Stockholm, Sweden}\\
       \email{sahari@kth.se}
\alignauthor Hadi Ghauch\\
       \affaddr{KTH Royal Institute of Technology,}\\
       \affaddr{School of Electrical Engineering,}\\
       \affaddr{Stockholm, Sweden}\\
       \email{ghauch@kth.se}
\alignauthor M. Mahboob Ur Rahman\titlenote{M. Mahboob Ur Rahman was affiliated with KTH from November 2013 - April 2016. Currently, he is an Assistant Professor at Information Technology University (ITU), Lahore, Pakistan.}
       \affaddr{KTH Royal Institute of Technology,}\\
       \affaddr{School of Electrical Engineering,}\\
       \affaddr{Stockholm, Sweden}\\
       \email{mahboob.rahman@ee.kth.se}
\and  
\alignauthor George Koudouridis\\
       \affaddr{Radio Network Technology Research,}\\
       \affaddr{Huawei Technologies,}\\
       \affaddr{Kista, Sweden}\\
       \email{george.koudouridis\\@huawei.com}
\alignauthor James Gross\\
       \affaddr{KTH Royal Institute of Technology,}\\
       \affaddr{School of Electrical Engineering,}\\
       \affaddr{Stockholm, Sweden}\\
       \email{james.gross@ee.kth.se}
}

\maketitle
\begin{abstract}
Explosive growth in the use of smart wireless devices has necessitated the provision of higher data rates and always-on connectivity, which are the main motivators for designing the fifth generation (5G) systems. To achieve higher system efficiency, massive antenna deployment with tight coordination is one potential strategy for designing 5G systems, but has two types of associated system overhead. First is the synchronization overhead, which can be reduced by implementing a cloud radio access network (CRAN)-based architecture design, that separates the baseband processing and radio access functionality to achieve better system synchronization. Second is the overhead for acquiring channel state information (CSI) of the users present in the system, which, however, increases tremendously when instantaneous CSI is used to serve high-mobility users. To serve a large number of users, a CRAN system with a dense deployment of remote radio heads (RRHs) is considered, such that each user has a line-of-sight (LOS) link with the corresponding RRH. Since, the trajectory of movement for high-mobility users is predictable, therefore, fairly accurate position estimates for those users can be obtained, and can be used for resource allocation to serve the considered users. The resource allocation is dependent upon various correlated system parameters, and these correlations can be learned using well-known \emph{machine learning} algorithms. This paper proposes a novel \emph{learning-based resource allocation scheme} for time division duplex (TDD) based 5G CRAN systems with dense RRH deployment, by using only the users' position estimates for resource allocation, thus avoiding the need for CSI acquisition. Also, an overhead model based on the proposed frame structure for 5G TDD is presented, both for user's position and its CSI acquisition. The proposed scheme achieves about 86\% of the optimal system performance, with an overhead of 2.4\%, compared to the traditional CSI-based resource allocation scheme which has an overhead of about 19\%. The proposed scheme is also fairly robust to changes in the propagation environment with a maximum performance loss of 5\% when either the scatterers' density or the shadowing effect varies. Avoiding the need for CSI acquisition reduces the overall system overhead significantly, while still achieving near-optimal system performance, and thus, better system efficiency is achieved at reduced cost.  
\end{abstract}

%
%
\printccsdesc



\section{Introduction}
Increased usage of smart electronic devices, such as hand-held mobile sets, tablets and laptops, in the recent years, has resulted in increased demand for higher data rates. Furthermore, the users of such devices demand full-time access to data packet connection, irrespective of their location and surrounding environment. Therefore, future communication systems are expected to have greater system efficiency and better provision of data service to the users compared to existing fourth generation (4G) technology \cite{5GU}. In the last few years, extensive research has been on going for the development and standardization of the fifth generation (5G) systems, that are expected to fulfil all the aforementioned requirements. Specifically, 5G systems will be able to provide a $\times$1000 increase in the system capacity \cite{5G_NetworkCapacity}, as well as almost $\times$10 decrease in latency \cite{Petteri_IEEE_Access}, compared to Long Term Evolution-Advanced (LTE-A) systems. Moreover, they will be able to provide high system efficiency and always-on connectivity, specially to high mobility users, in Ultra-Dense Network (UDN) deployments \cite{Hesse_PIMRC}. 

To achieve these goals for 5G, one possible approach is to massively increase the number of antennas (either centrally or de-centrally) \cite{C_and_D_MIMO}. Research from the last few years indicates that significant performance gain can be obtained from massive antenna deployment, if transmission from such antennas is tightly coordinated \cite{IA_MIMO}, \cite{Rate_MIMO}. This tight coordination includes phase-level synchronization, which is needed for joint transmission, as well as the synchronization needed for coordinated pre-coding. Using tight synchronization between these large number of antennas leads to a coordination overhead, as discussed in \cite{Sync_OH_SPAWC2014}. To overcome this problem, the cloud radio access network (CRAN) architecture has been introduced, which is a centralized, cloud-computing based network architecture \cite{5G_CRAN}. In CRAN, the baseband units (the main signal processing units of the network) are connected to the cloud to form a pool of centralized processors, which is then connected to the set of distributed antennas (the radio access units) in the system. Thus, separating the baseband units from the radio access units helps in achieving synchronized coordination between large sets of antennas, at a relatively reduced cost in the system. However, besides the synchronization overhead, the overhead for acquiring channel state information (CSI) of the mobile users still exists, which increases with the number of antennas, the granularity of the CSI to be acquired as well as the mobility of the terminal users. For achieving the aforementioned system requirements of 5G, the cost of acquiring CSI has to be minimized, which is the main topic addressed in this paper.

The main purpose of CSI acquisition is to perform allocation of resources such that all users can be served well. The resources include time and/or frequency resources, coding rates, modulation schemes, transmit beamforming, and many more. Much work has been done in the past few years for designing efficient resource allocation schemes, specific to certain 5G system characteristics. A non-orthogonal resource allocation scheme, called non-orthogonal multiple access (NOMA) \cite{5G_Tech_dir}, has been investigated in \cite{NOMA}, for increased system throughput and accommodating maximum users by sharing time and frequency resources. The technique of dynamic time domain duplexing for centralized and decentralized resource allocation in 5G has been studied in \cite{Hesse_PIMRC}. In \cite{Multi_beam_op}, a radio resource allocation scheme for multi-beam operating systems has been proposed, where the radio resources are allocated to a user based on its channel state and the resources within the beam serving that user. The authors in \cite{ICT_2015_Rostami} propose a resource block (RB) allocation algorithm, which exploits the combination of multi-user diversity and users' CSI for allocation of RBs, with carrier aggregation, and modulation and coding scheme (MCS) for throughput maximization in 5G LTE-A network. 

Some of these resource allocation schemes exploit the users' CSI, which incurs a significant system overhead for high mobility users, but this overhead is not considered in those studies. On the other hand, the system's performance is affected if outdated CSI is used for resource allocation for high mobility users \cite{Shen_massiveMIMO}. One of the network deployment architectures suited for achieving the expected targets of a 5G system is the ultra-dense small cell deployment, in which the users are expected to be in line-of-sight (LOS) with the serving base stations at almost all times. In this case, the users' position information can be used instead of their CSI \cite{my_paper_PIMRC}. Essentially, the optimal allocation of resources is dependent upon the system parameters (including users' position, users' velocity, propagation environment, interference in the system, and so on), which are inherently correlated. One way of exploiting these hidden correlations among system parameters for efficient allocation of resources is through machine learning, which is the method proposed in this paper. Previously, various machine learning algorithms have been used for resource allocation in different domains of wireless communication systems; some examples include using support vector machines (SVMs) for power control in CDMA systems \cite{IEEE_2003_J_A_Rohwer}, prediction of the next cell of a mobile user using supervised learning techniques and CSI \cite{IEEE_2013_SPAWC}, rate adaptation using random forests (a form of supervised machine learning technique) in vehicular networks \cite{IEEE_2013_Oscar}, and many more. Use of machine learning has also been investigated for orthogonal frequency division multiplex-multiple input, multiple output (OFDM-MIMO) based 5G systems \cite{Joao_WCNC_2012}, \cite{Alvarino_IEEEWC_2014}. However, for time division duplex (TDD) MIMO systems, the resource allocation is done based on instantaneous CSI availability (without using learning, or considering the CSI acquistion overhead), where resource allocation is referred to RB assignment \cite{ICT_2015_Rostami}, rate allocation \cite{Tassiulas_HCN_IEEE_VT_2015} and beamforming for joint transmission-based coordinated multipoint (CoMP) \cite{Wang_TDD_Het_VT2013}. 

This paper discusses the use of machine learning for designing a novel learning-based resource allocation scheme for TDD multi-user MIMO (MU-MIMO) CRAN systems based on the position estimates of high-mobility users, without using instantaneous CSI. For this purpose, `random forest' algorithm is used, and resources including transmit beam, receive filter and packet sizes are assigned to the intended users based on their position estimates (which can be accurate or inaccurate). The robustness of the proposed resource allocation scheme is tested by using different values in training and test datasets for random forest, such as using accurate position estimates for training the random forest and testing its performance using data having inaccurate position estimates of the users. Afterwards, the system goodput is computed for the proposed resource allocation scheme and is compared to the system goodput when instantaneous CSI of users (with a system overhead) is used for resource allocation. The results show that the proposed scheme achieves about 86\% of the system performance obtained for traditional CSI-based resource allocation scheme. Furthermore, a maximum performance loss of 5\% is observed when either the scatterers' density or the shadowing effect varies, thus showing the robustness of the proposed scheme to changes in the propagation environment. 

To highlight the effectiveness of the proposed scheme, an overhead model based on the frame structure for 5G TDD proposed in \cite{ICC_2016_Petteri} is also presented, for both the user's position and CSI acquisition, and its effect on the system throughput is evaluated. The results show that the proposed scheme, which is based on user's position acquisition, incurs a system overhead of only 2.4\% compared to the traditional CSI acquisition-based resource allocation which has an overhead of 19\%. The structure for the rest of the paper is as follows: Section \ref{sys_model} presents the system model, as well the details regarding the overhead model for 5G TDD. Details of the proposed learning-based resource allocation scheme are presented in section \ref{MbRA}, along with a brief background on machine learning algorithm `\emph{random forest}'. Simulation results and relevant discussions are presented in section \ref{RnD}. Section \ref{con} concludes the paper.

\section{System Model}
\label{sys_model}
\begin{figure}[!]
\centering
\includegraphics[width=5cm]{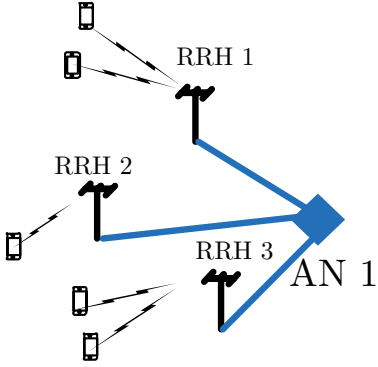}
\caption{The CRAN architecture for 5G system}
\label{CRAN}
\end{figure}

Consider a scenario (Figure \ref{CRAN}), based on CRAN architecture, where $N$ users are being served by $R$ remote radio heads (RRHs), and all RRHs are connected to an aggregation node (AN). The AN is the computational hub where all baseband processing takes place, whereas RRHs mainly serve as radio frequency (RF) front ends. Further details of the CRAN based system model can be found in \cite{Our_paper_GC}. This work focuses on the downlink communication of the aforementioned 5G CRAN system model. A TDD based frame structure is considered for downlink communication, and the operational frequency of the CRAN system is $f_c$. The users are assumed to be moving at high speeds ($v_{\text{Rx}} > 50$ km/h). The RRHs are densely deployed (UDN deployment), such that the users are expected to be in LOS with the serving RRHs. Also, each user is equipped with $N_{\text{Rx}}$ antennas, each at a height of $h_{\text{Rx}}$ from the ground, and will be served by an RRH having $R_{\text{Tx}}$ antennas, each at a height of $h_{\text{Tx}}$ from the ground. 

The channel between the RRH $r$ and user $n$ is characterized by the spatial system parameters [such as the angle of arrival (AoA) and the angle of departure (AoD)], the frequency-based system parameters (such as operational frequency of the system, and the Doppler shift), as well as the time-dependent system parameters (such as change in user's position, change in scatterers' density, propagation environment, etc.). All RRHs are expected to serve at least one user, in the same time-slot, implying that all users will experience interference from other users being served by the same RRH, as well as cross-channel interference from the neighboring RRHs. Each RRH is connected to the AN, which acts as a resource allocation unit, and consists of a set of resources, including transmit beams, receive filters, and packet sizes, to serve a given user. Full-buffer condition is assumed, which means that at each time, at least one user $n$ needs to be allocated resources by the AN for being served by the associated RRH $r$. A fixed set of transmit beams $B_{\text{Tx}}$ is available to serve the users, based on the geometry of the propagation scenario, and is also used to design a set of receive filters $B_{\text{Rx}}$, which will be used by the terminal users for reduced-interference reception. The position coordinates $\mathcal{P}_{n,(x,y,z)}$ of the $n^{th}$ user are available at the $r^{th}$ RRH, with some inaccuracy, and is the primary parameter used for allocation of resources by the AN connected to the RRH.   

For simplifying the analysis, we consider that each RRH is serving only one user in a given time-slot, so that only cross-channel interference exists in the system. Based on all these parameters, the signal-to-interference-and-noise ratio (SINR) for a user $n$, at time $t$, is calculated as follows:
\begin{align}
\label{SINR_equation}
\gamma_{n, t}(\phi_n^a, \phi_n^d) = \frac{ P_{n, t}(\phi_n^a, \phi_n^d)  }{ \sigma^2 + \sum\limits_{ \substack{m=1 \\ m \neq n } }^{N} P_{m, t}(\phi_n^a, \phi_m^d) },
\end{align}

where, $P_{n, t}$ is the received signal power for a user $n$, at time $t$, and is given by:
\begin{align}
\label{received_power_equation}
\begin{split}
P_{n, t}(\phi_n^a, \phi_n^d) = & \ P_{\text{Tx}} h_{\text{PL}}^2 \cdot \\ 
& \cdot \ | \pmb{U}(\phi_n^a)^\dagger \pmb{H}_t(\phi_n^a, \phi_n^d) \pmb{V}( \phi_n^d )|^2 .
\end{split}
\end{align}

Here, $P_{\text{Tx}}$ is the allocated transmit power, $h_{\text{PL}}^2$ denotes the pathloss, $\phi_n^a$ is the azimuth AoA of user $n$, and $\phi_n^d$ is its azimuth AoD. $\pmb{U}(\phi_n^a)$ is the receive filter with the main beam focused in the direction closest to $\phi_n^a$, and $\pmb{V}(\phi_n^d)$ is the transmit beamformer with the main beam located in the direction closest to $\phi_n^d$ (details regarding beamforming can be found in \cite{molisch2007wireless}). $\pmb{H}_t(\phi_n^a, \phi_n^d)$ is the channel matrix for an instance of time $t$ for a given $\phi_n^a$ and $\phi_n^d$, and $\sigma^2$ is the noise power. $(.)^\dagger$ denotes the Hermitian of a vector or a matrix.

The SINR computed for a given combination of $\pmb{U}(\phi_n^a)$ and $\pmb{V}(\phi_n^d)$, with the corresponding channel matrix $\pmb{H}_t(\phi_n^a, \phi_n^d)$, is used to compute the transport capacity for user $n$ by the following formula:
\begin{align}
\label{trans_cap}
C_{n, t} = S \times \log_2(1 + \gamma_{n, t}(\phi_n^a, \phi_n^d)).
\end{align}

Here, $S$ is the symbol length, which is the product of the transmission time interval (TTI) and bandwidth $BW$ of the system. For determining the transmission success or failure, the error model based on Shannon's capacity (Eq. (\ref{trans_cap})) is used; if the $n^{th}$ user's packet size $ < C_{n, t}$ then the packet is successfully transmitted, otherwise the packet transmission for user $n$ fails. 

\subsection{The Overhead Model}
\label{OH_model}
The frame structure proposed in \cite{ICC_2016_Petteri} for 5G TDD based system is considered for formulating an overhead model. The total length of the frame is 0.2 ms and it consists of 42 OFDM symbols ($T_{\text{sym,total}}$ = 42), and about 833 sub-carriers ($ f_{\text{sc,total}}$ = 833). The position information of the users present in the system can be acquired using narrow-band pilots (also called beacons), typically spanning the first symbol of the frame. The CSI for the users can be obtained using 4 full-band pilots, placed at the beginning of a frame just after the positioning beacons. The adjacent CSI-sensing pilots are scheduled based on the cyclic-prefix compensation distance, as explained in \cite{Petteri_IEEE_Access}, to avoid inter-carrier interference. Based on these parameters, the overhead for position acquisition per user can be calculated as:

\begin{align}
\label{OH_pos}
OH_{pos,n} = \frac{T_{\text{sym,pos,n}} \times f_{\text{sc,pos,n}} }{T_{\text{sym,total}} \times f_{\text{sc,total}}}
\end{align}

Here, $T_{\text{sym,pos,n}}$ is the number of OFDM symbols used for position estimation of user $n$, and $f_{\text{sc,pos,n}}$ denotes the number of sub-carriers used in the positioning beacon. Similarly, for CSI acquisition per user, the overhead can be computed as:
\begin{align}
\label{OH_CSI}
OH_{CSI,n} = \frac{T_{\text{sym,CSI,n}} \times f_{\text{sc,CSI,n}} }{T_{\text{sym,total}} \times f_{\text{sc,total}}},
\end{align}

where $T_{\text{sym,CSI,n}}$ and $f_{\text{sc,CSI,n}}$ denote the number of OFDM symbols and the number of sub-carriers, used for CSI acquisition of user $n$, respectively. The system overhead for position, or CSI, acquisition related resource allocation scheme can be computed by multiplying the corresponding overhead with the number of users for which the position information, or CSI, is acquired. 

\subsection{Problem Statement}
In the considered CRAN system, the task of the AN is to allocate the resources efficiently for each RRH-user link, per TTI, such that the system's sum-throughput is maximized. For this purpose, it needs to acquire the CSI of all users in the system, on per TTI basis, which incurs a large system overhead. The task of efficient resource allocation becomes further challenging for high-mobility users particularly, where CSI acquisition is crucial for achieving maximum sum-throughput of the system. 

One way of avoiding the CSI acquisition overhead is to use the position information of the high-speed users; since LOS exists, the resource allocation for users can be done based on their position information rather than using their instantaneous CSI. However, this position information can not be used directly for efficient resource allocation, rather, the hidden correlation among the position estimates and the other system parameters has to be exploited together for this purpose. We propose to use \emph{machine learning} for accomplishing this task. Specifically, we use machine learning to design a resource allocation scheme for the aforementioned system, purely based on the users' position information, such that the CSI acquisition is not needed at all. We will investigate the performance of this \emph{learning-based resource allocation scheme} in comparison to the conventional resource allocation technique, where CSI acquisition is needed, also taking into account the system overhead. Furthermore, we want to test the robustness of the learning-based resource allocation scheme, when the position information for the users in the system is inaccurate. In the next section, we discuss the details regarding the design of the learning-based resource allocation scheme, along with some background on machine learning.

\section{Design of the Learning-based Resource Allocation Scheme}
\label{MbRA}
Learning the different correlated parameters is accomplished using \emph{machine learning}, which is defined as ``the capability of a computer program or a machine to develop new knowledge or skill from the existing and non-existing data samples to optimize some performance criteria" \cite{alpaydin2014introduction}. `Random forest algorithm' \cite{Breiman_RF} is the learning technique used in this work for learning the system parameters, and predicting the probability of successful or failed transmission of data packets from a given RRH to the respective user(s). We first provide some background on the random forest algorithm, followed by the details about how can this algorithm be used for designing the learning-based resource allocation scheme.

\subsection{Background on Random Forest Algorithm}
\label{RF}
\begin{figure}
\centering
\includegraphics[width=8cm]{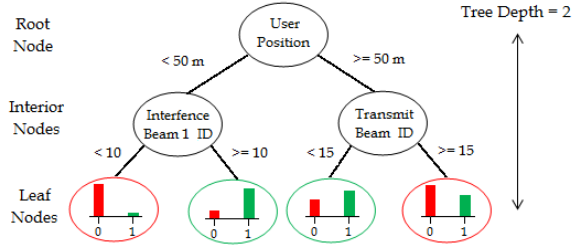}
\caption{An example of a binary random decision tree}
\label{fig_my_decision_tree}
\end{figure}

Random forest algorithm is a supervised learning technique, which consists of a number of random decision trees (hence the term `forest') that are built, using the statistical information of the supplied dataset, to develop a hypothesis for predicting the outcome of a future instance \cite{IEEE_2013_Oscar}, \cite{ref_1_IEEE_2012_A_Chaudhry}. Each instance $x$ of the dataset $\pmb{x}$ consists of two parts: a set of data characteristics $\pmb{I}$, called \emph{features}, and the relevant output variable $y$, and collectively they form the \emph{input feature vector} $x_i$. To learn the information in the data features $\pmb{I}$ (the `training' process), the random forest algorithm constructs $T_n$ binary random decision trees, each having a maximum depth $T_d$. Each tree has one \emph{root node} and several \emph{interior} and \emph{leaf nodes}. Figure \ref{fig_my_decision_tree} shows an example of a binary random decision tree, having some interior and leaf nodes. The classification features of the decision variable are taken from the input feature vector, and are represented by the root and interior nodes in the random decision tree. Each root node and interior node is constructed by a decision threshold based on a (randomly selected) feature subset from the set of given $\pmb{I}$ input features. Thus, each tree has a different subset of features considered for decision threshold at each of its nodes. The output variable is represented by the leaf nodes of a decision tree. The instance on which the prediction has to be made, is traversed through all decision trees in the forest, to get $T_n$ output variables, called \emph{votes}. The output variable $y$ is predicted by aggregating all the votes and selecting the majority class (category or value of the output variable) from among those votes. 

For making each tree in the training phase, a training dataset $\pmb{z}$, having the same size as the training data $\pmb{x}$, is constructed by using training samples which are randomly chosen, with replacement, from $\pmb{x}$. This random selection with replacement makes some instances from the training data to be used repeatedly, while some are not used at all. The later instances are collectively known as \emph{out-of-the-bag} (OOB) examples and represent almost 30\% of the total training data \cite{Breiman_RF}. A random subset of input variables is used for every node of a decision tree from the $\pmb{z}$ training examples. A decision threshold is determined for the selected input variable, based on which the left or right traversing path in the subsequent levels of the random decision tree is chosen. It is critical to select the input variable at a node, as well as the decision threshold, such that the \emph{purity} of the subsequent levels of the random decision tree is maximized. Purity measures the extent to which the resulting child node is made up of cases having the same output variable \cite{oracle_doc}. Thus, an ideal threshold at any node would divide the data in such a way that the resulting child nodes would give distinct values of the output variable. 

The generated random forest has two types of qualitative measures. First is the \emph{prediction accuracy}, which measures how accurately the random forest predicts the output variable for a given dataset. If the prediction accuracy is evaluated on the training data, it is called the \emph{training accuracy}, while the same when evaluated on a newly collected dataset is called the \emph{test accuracy}. Second qualitative measure is the \emph{importance of an input variable}, which indicates how important is a particular input variable in determining the desired output variable. In general, the random forest algorithm can cater for the missing input data variables, is robust to noisy data and is computationally efficient \cite{Breiman_RF}. Also, it does not suffer from the problem of over-fitting, by using only a subset of the training data for making the random decision trees which make up the random forest. Due to all these properties, the random forest algorithm has been previously used in designing different techniques for optimal system performance. Some examples include using random forest algorithm for intrusion detection for mobile devices \cite{ref_19_IEEE_2012_A_Chaudhry}, and designing a rate adaptation scheme in vehicular networks using the random forest \cite{IEEE_2013_Oscar}. 
 
\subsection{The Learning-Based Resource Allocation Scheme}
\label{LbRA}
The main aim of the learning-based resource allocation scheme is to use only the position estimate of the users and learn its relationship with different system parameters and resources, such that the system resources are efficiently utilized without incurring excessive overhead. We first explain the structure of the learning-based resource allocation scheme, and then present its working details. 

\subsubsection{Structure of the Learning-Based Resource Allocation Scheme}
\label{S_LbRA}
The structure of the learning-based resource allocation scheme can be divided into three parts: the pre-processing unit, the machine learning unit, and the scheduler. 

\subsubsection*{The Pre-Processing Unit}
The pre-processing unit plays an important role in the training of the machine learning unit, by helping in designing the training dataset. The training dataset is constructed off-line, and hence the CSI as well as the position information of the users are available at the AN, along with the information for the resources to be allocated. In (off-line phase of) the pre-processing unit, the optimal transport capacity for each user is computed using its CSI (considering all the other users in the system), based on the maximization of the system's sum-transport capacity. Then, the optimal transmit beam $b_{\text{Tx}}$ and receive filter $b_{\text{Rx}}$ combinations for a given user position $\mathcal{P}_{n,(x,y,z)}$ are identified, for which the optimal transport capacity is obtained (i.e. the exhaustive search), and are used as the input features for the training dataset of the machine learning unit. Based on the values of the optimal transport capacity for the overall system, a set of packet sizes is designed, which consists of 5 discrete values, and the optimal transport capacity for each user is checked against those packet sizes (according to the Shannon's capacity-based error model) to generate the output variables, 0 or 1, for the training dataset. Thus, the user's ID $n$, its position information $\mathcal{P}_{n,(x,y,z)}$, optimal transmit beam $b_{\text{Tx}}$, optimal receive filter $b_{\text{Rx}}$, the packet size $PS$, and the output variable (0 or 1) form the input feature vector, and a set of those input feature vectors makes up the training dataset to be used by the machine learning unit.

\subsubsection*{The Machine Learning Unit} 
The training of the machine learning unit is done off-line, where the training dataset is used to learn the input features, i.e. the user's ID $n$, its position information $\mathcal{P}_{n,(x,y,z)}$, optimal transmit beam $b_{\text{Tx}}$, optimal receive filter $b_{\text{Rx}}$, and the packet size $PS$. The learning is essentially done to construct the random forest, with the parameters like number of decision trees $T_n$, tree depth $T_d$ and number of random features for split at each tree node, chosen so as to optimize the training accuracy of the random forest. Here, it should be noted that the performance of the random forest is affected by the `bias' in output variable distribution for the overall training dataset, i.e. the training accuracy is affected if, for example, a large number ($> 80\%$) of input feature vectors have class `0' as output variable than class `1', and vice versa. This bias in class distribution is being taken care of by the pre-processing unit, such that the training dataset has a balanced number of input feature vectors for both the classes `0' and `1', as the output variable. Once an optimal training accuracy is achieved, the machine learning unit is ready to be used for testing new dataset(s) generated on run-time in a realistic system.

\subsubsection*{The Scheduler}
In a realistic system, the scheduler is the main component responsible for forwarding the information about the allocated resources for all users in the system. This proposed scheme includes a scheduler as the last unit, whose main task is to forward the information about the allocated resources (obtained from the machine learning unit) for each user in the system, to their corresponding RRH. This scheduler is, however, sensitive to the occurrence of false-positives in the output from the machine learning unit. Technically, a false-positive occurs when an input feature vector has `0' as its output variable realistically, but the learning algorithm wrongly predicts the output variable to be `1' for that input feature vector. In the proposed scheme, false-positive occurrence makes the algorithm more error-prone, by suggesting a higher packet size $PS_{o+1}$ to serve a particular user, though, realistically, the highest packet size that can serve the user is $PS_o$. In this case, the scheduler backs-off the packet size for transmission, and transmits a packet size, chosen randomly, from the set of packet sizes one less than $PS_{o+1}$, i.e. the packet size for which the false positive detection occurred. We call this a `random back-off scheduler', which operates in combination with the output predicted by the random forest, and thus completes the design structure of learning-based resource allocation scheme. The false-positive occurrence is identified from the output variables available in the training dataset, and based on this, the scheduler operates more sensitively for those input feature vectors. In this way, the resource allocation scheme ensures that erroneous working of the machine learning unit does not significantly impact the system's performance.

\begin{figure}[!]
\centering
\includegraphics[width=8cm]{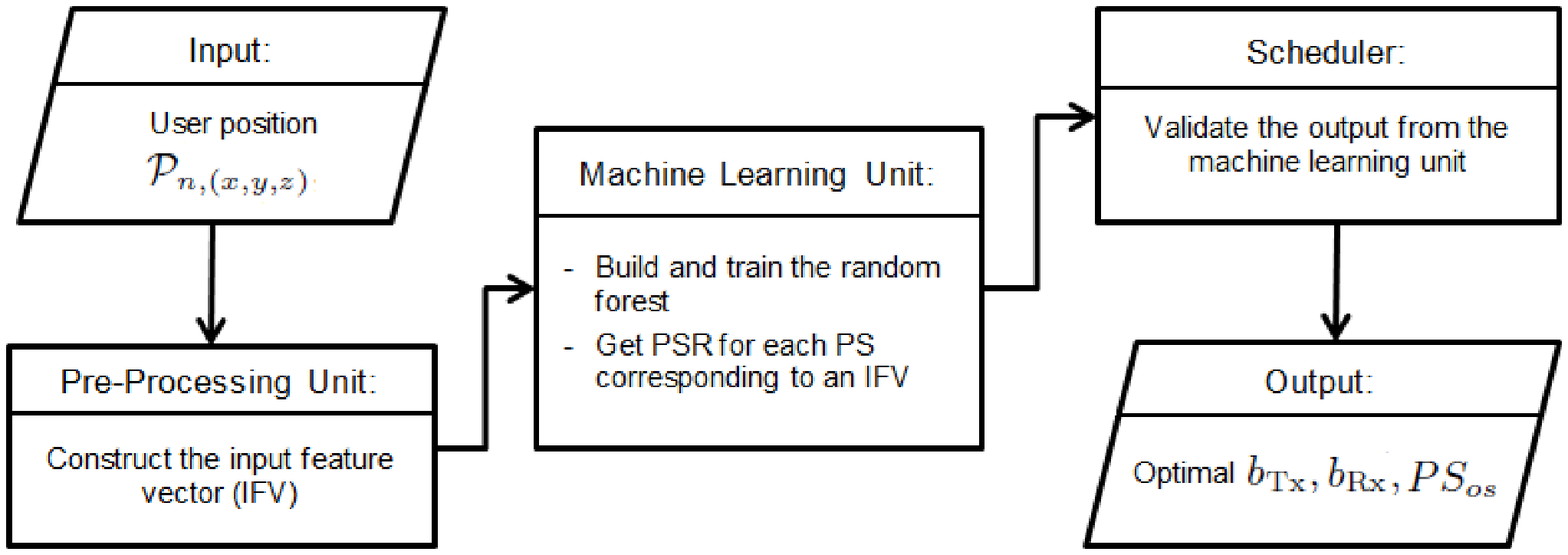}
\caption{The proposed Learning-Based Resource Allocation Scheme}
\label{fig_LbRA}
\end{figure}

\subsubsection{Working of the Learning-Based Resource Allocation Scheme}
\label{W_LbRA}
In a realistic system, the position estimate $\hat{\mathcal{P}}_{n,(x,y,z)}$ of the user $n$ is acquired by the corresponding RRH and reported back to the AN. This position estimate is used by the pre-processing unit, where it is compared against the user position information $\mathcal{P}_{n,(x,y,z)}$ available in the training dataset, and the position information in the training data that gives the minimum value for $|\mathcal{P}_{n,(x,y,z)} - \hat{\mathcal{P}}_{n,(x,y,z)}| $ is chosen to construct the input feature vector for the test dataset. Once the closest position estimate $\mathcal{P}_{n,(x,y,z)} ^o$ is obtained, it is combined with the corresponding optimal transmit beam $b_{\text{Tx}}$, receive filter $b_{\text{Rx}}$, and with the 5 discrete packet sizes $PS$ to form a set of input feature vectors for different packet sizes corresponding to the position estimate $\hat{\mathcal{P}}_{n,(x,y,z)}$. 

This set of input feature vectors is then passed to the machine learning unit, where each of the input feature vector is parsed through the random forest to obtain the \emph{votes} for the predicted output variable by each decision tree in the forest. In essence, the votes are obtained for successful transmission (i.e for $y = 1$) of a packet size $PS_p$ and denote the packet success rate (PSR) for $PS_p$. This PSR also denotes the tendency of the machine learning unit's predicted output variable; if the PSR $\geq T_n/ 2$, then the predicted output variable is `1', otherwise, it is `0'. This predicted output variable is tested for false-positive detection by the scheduler, by comparing it to the output variable for the corresponding input feature vector in the training dataset, which then, either retains the packet size $PS_p$ if the prediction is correct, or chooses a random packet size $PS_r$ in case of false-positive occurrence, to give the optimal packet size $PS_{os}$ predicted for transmission by the learning-based resource allocation scheme. The PSR corresponding to $PS_{os}$ is used to compute the system goodput predicted by the learning-based resource allocation scheme, as follows:
\begin{align}
Goodput_{os} = PSR_{os} \times PS_{os}
\end{align}

The optimal transmit beam $b_{\text{Tx}}$, receive filter $b_{\text{Rx}}$ and packet size $PS_{os}$ predicted by the learning-based resource allocation scheme is considered to belong to that set of instances for all users which achieves the maximum sum-goodput. Figure \ref{fig_LbRA} shows the different steps of the proposed learning-based resource allocation scheme. Overall, the random forest algorithm is expected to learn the assignment of optimal packet size, transmit beam and receive filter for each user, in order to maximize the system goodput, using only the users' position information but without knowing their CSI. In reality, the position estimates of high-mobility users can be acquired with certain precision using an extended Kalman filter (EKF), along with the direction of arrival (DoA) and time of arrival (ToA) estimates of those users \cite{EKF_Mario}. This means that it is not possible to always have the accurate position information for the users in the system. Since the random forest algorithm is robust to noisy data, the learning-based resource allocation scheme is expected to perform well when noisy position estimates of the users are available for either the training or test datasets (or both). Once the proposed scheme suggests the resources $b_{\text{Tx}}$, $b_{\text{Rx}}$, and $PS_{os}$ for serving a given user $n$, this information is passed on the corresponding RRH $r$, which further sends a pilot signal to the user $n$ to inform it regarding the receive filter $b_{\text{Rx}}$, suggested by the proposed scheme, for reduced-interference reception. The performance of the proposed resource allocation scheme is tested by performing system-level simulations, the details of which are given in the next section, along with the results and related discussions.

\section{Results and Discussion}
\label{RnD}
In this section, we first compare the performance of the proposed scheme to that of the traditional resource allocation scheme based on user's CSI. We also investigate the robustness of the proposed scheme when inaccurate position estimates are available in the test dataset, or when the propagation environment parameters vary in the training and the test datasets (specifically, change of scatterers' density and change in shadowing characteristics). Afterwards, we present the effect of overhead on the proposed and the traditional schemes on the theoretical system throughput for a 5G CRAN system.

\subsection{Evaluation Methodology}
\begin{figure}[!]
\centering
\includegraphics[width=7cm]{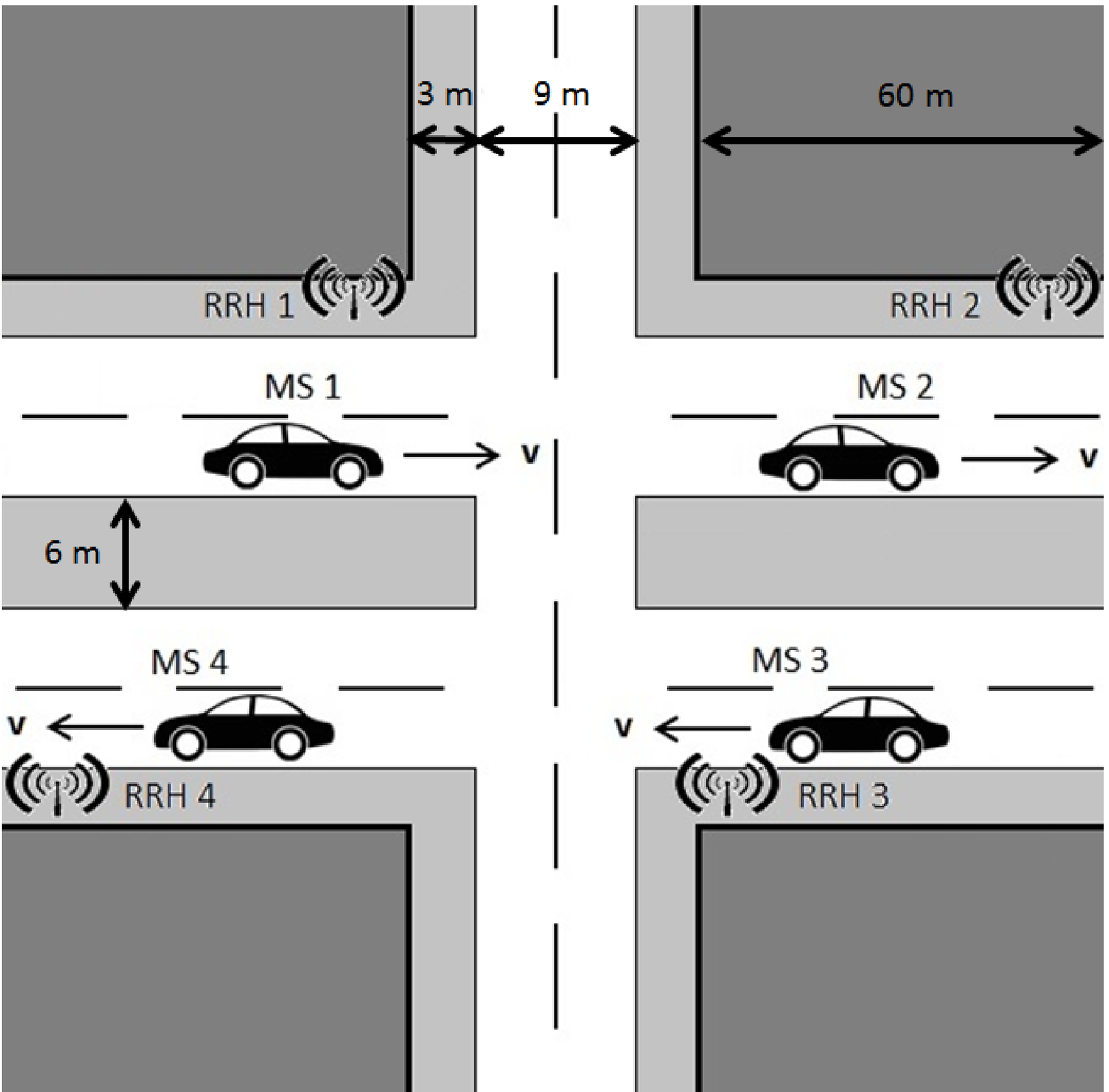}
\caption{The simulation scenario; each RRH serves one user}
\label{sim_scenario}
\end{figure}

The performance evaluation of the proposed scheme in section \ref{LbRA} is done by doing realistic simulations using a discrete event simulator (DES) called Horizon \cite{ref_29_D1_SofA}. The propagation scenario, shown in figure \ref{sim_scenario}, is implemented in Horizon for simulating a CRAN based multi-users, multi-RRHs communication system, as presented in section \ref{sys_model}. Based on the propagation scenario, a fixed set of transmit beams is designed in the following way: the transmit beams are formed using geometric beamforming, where each consecutive beam is separated by 3$^{\circ}$ angular resolution. The receive filters are, essentially, geometric beams formed by using the multiple antennas at the user end, and are designed in the same way as the transmit beams but with an angular resolution of 12$^{\circ}$. The parametrization for system simulations is given in table \ref{table_sim}. The channel coefficients for downlink communication are extracted from the simulator for each TTI, i.e. after every $1$ ms. Ray-tracing based METIS channel model \cite{METIS_D1.2} for Madrid grid is implemented in Horizon for generating the channel coefficients. Details about the ray-tracer based channel model can be found in \cite{Location_BF_UDN}.

\begin{table}
\caption{Parameter Settings for the Simulator}
\label{table_sim}
\centering
\begin{tabular}{c c}
\hline
Parameter & Value\\
\hline
$f_c$ & 3.5 GHz\\
$BW$ & 5 MHz\\
$R_{\text{Tx}}$ & 8\\
$N_{\text{Rx}}$ & 2\\
$h_{\text{Tx}}$ & 10 m\\
$h_{\text{Rx}}$ & 1.5 m\\
$P_{\text{Tx}}$ & 1 mW\\
$TTI$ & 1 ms\\
$v_{Rx}$ & 30 m/s\\
\hline
\end{tabular}
\end{table}

After computing the channel matrices, the training dataset is generated using the procedure explained in section \ref{S_LbRA}. As mentioned earlier, the training data is used to build random forests for various parameter settings, from which the random forest with the optimal training accuracy is chosen for further processing. The random forest is constructed using the random forest implementation in WEKA software \cite{hall2009weka}. Table \ref{table_RF} shows the values of training accuracy obtained for different parameter settings of random forest algorithm. Based on the results, the random forest with $T_n = 10$ and $T_d = 3$ was chosen for further processing, with the number of random features used for split at each node of decision tree as $\sqrt{I}$ \cite{Breiman_RF}. Here, it should be noted that selecting the random forest with the highest training accuracy (in our case, for $T_n = 20$ and $T_d = 4$) is not always the best choice; having a larger number of trees for a small set of input features $\pmb{I}$ increases the correlation among the trees (thus reducing the robustness of the random forest to noisy data), as well as increases the computation time for constructing the random forest.

\begin{table}
\renewcommand{\arraystretch}{1.3}
\caption{Training Accuracy of Random Forest for Different Parameter Settings}
\label{table_RF}
\centering
\begin{tabular}{c c c}
\hline
\multicolumn{1}{c}{$T_n$} & \multicolumn{1}{c}{$T_d$} & Training Accuracy (\%)\\
\hline
5 & 3 & 83.3\\
\textbf{10} & \textbf{3} & \textbf{86}\\
10 & 4 & 86.9\\
20 & 3 & 86.65\\
20 & 4 & 87.2\\
\hline
\end{tabular}
\end{table} 

A total of 100 user positions (for each user) are chosen randomly from the available set of 1000 positions (for each user) in the overall simulation scenario, for generating the training and test datasets, each having 0.25 million samples. The output from the random forest is combined with the scheduler, as explained in section \ref{W_LbRA}, and the system goodput (in bits/TTI) is computed. The first simulation is performed by setting the scattering objects' density as $0.01$/m$^2$, i.e., 1 scatterer per 10$\times$10 m$^2$ area. The performance of learning-based resource allocation scheme is compared against the following schemes:

\begin{itemize}
\item Random packet scheduler: Schedules a randomly selected packet size for each user using the optimal selection strategy for transmit beam and receive filter.
\item Random packet scheduler for geometric beam and filter assignment: Schedules a randomly selected packet size for each user using the location-based assignment of transmit beam and receive filter.
\item Optimal packet scheduler (the Genie): Schedules the optimal packet size for each user based on the optimal transport capacity for each user, obtained through the instantaneous CSI of the users.
\end{itemize}

\subsection{Results for the Proposed Scheme}
\label{results}

\begin{figure}[!]
\centering
\includegraphics[width=9cm,height=5cm]{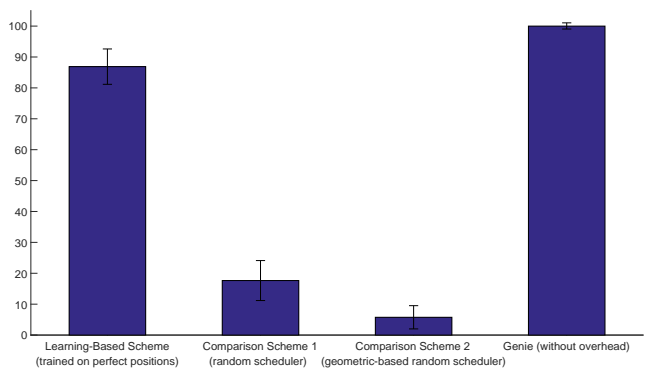}
\caption{System goodput (in \%age relative to the Genie) for perfect users' positions information}
\label{results_perfect_pos}
\end{figure}

\begin{figure}[!]
\centering
\includegraphics[width=9cm,height=5cm,keepaspectratio]{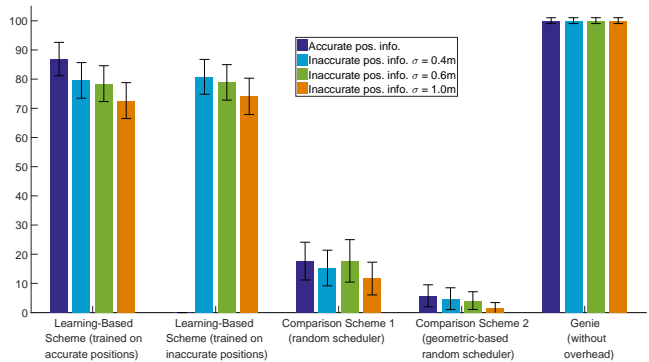}
\caption{System goodput (in \%age relative to the Genie) for different schemes and various possibilities of available position information}
\label{results_all_pos}
\end{figure}

Figure \ref{results_perfect_pos} shows the system goodput obtained for each of the resource allocation schemes when perfect information about each user position is available. The results are shown as the system goodput relative to the one obtained by the Genie. It can be seen that the learning-based resource allocation scheme performs very well compared to the Genie, and achieves about 86\% of the optimal system performance (i.e. Genie without system overhead). The training accuracy of the random forest is 86\%, where the performance loss is due to the inequitable distribution of output variables in the training dataset. The random packet scheduler performs worse, which highlights the importance of learning the system parameters for optimal resource allocation. The geometric assignment-based random scheduler also shows poor performance (only 6\% goodput compared to the optimal one), due to the fact that geometric-based allocation of transmit beam and receive filter is not the optimal strategy for serving a user in an interference-limited system.

In reality, the perfect position information is not always available, rather there is some inaccuracy involved in the reported coordinates for the user's position. Figure \ref{results_all_pos} shows the relative system goodput for all resource allocation schemes when the user position is having an inaccuracy variance of 0.4, 0.6 and 1.0 m. It can be seen that the position inaccuracy affects the system performance for all sub-optimal resource allocation schemes due to the fact that optimal transmit beam and receive filter combination is not valid for the inaccurate position information. Despite this, the learning-based allocation scheme achieves more than 72\% of the optimal system performance (for the highest inaccuracy variance), which is 4 times better than any of the other comparison schemes. For having a fair performance comparison, we trained the random forests for each of the cases of inaccurate position availability, and tested their performance against the relative test data for inaccurate position information. The results show that no significant improvement in performance can be obtained if the learning is performed for inaccurate position information datasets; the random forest trained on accurate user position information can also operate effectively for any case of inaccurate user position information.

\begin{figure}
\centering
\includegraphics[width=8.5cm,height=5cm,keepaspectratio]{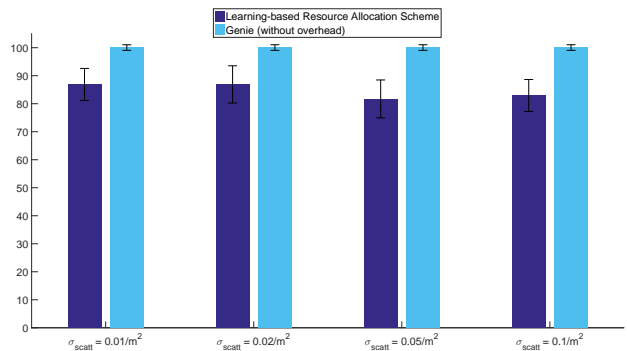}
\caption{System goodput (in \%age relative to the Genie) for different scattering densities for perfect users' position information}
\label{results_all_scatt}
\end{figure}

To observe the effect of randomness in the system parameters on the performance of different resource allocation schemes, the scatterers' density is varied. Figure \ref{results_all_scatt} shows the relative system goodput for learning-based resource allocation scheme for different values of scattering objects' density when perfect user position is available. The random forest in the machine learning unit is trained for scatterers' density of $0.01$/m$^2$ (the same as used for previous simulations), and is tested for datasets generated using different values of scatterers' density. The results show that the relative system goodput is not affected severely when learning-based resource allocation scheme is used for changing scatterers' density in the propagation environment. The maximum difference with respect to the Genie is 83\% (for 10 scatterers per 100 m$^2$ area), when the dataset generated for different densities of scattering objects is tested against the random forest generated using a fixed scatterers' density. Realistically, the goodput of the system is expected to be not affected severely by the change of scatterers' density, since LOS link exists at all times between the users and their corresponding RRHs. Keeping this into consideration, the learning-based resource allocation scheme is seen to be robust for changing scatterers' density, where the maximum performance loss compared to the Genie varies by less than 5\% as the number of scatterers per 100 m$^2$ of area is increased.

\begin{figure}
\centering
\includegraphics[width=8.5cm,height=5cm,keepaspectratio]{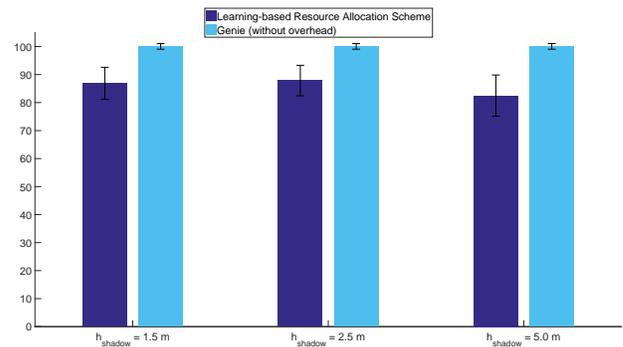}
\caption{System goodput (in \%age relative to the Genie) for different heights of the shadow object for perfect users' position information and $\sigma_\text{scatt}$ = 0.01/m$^2$}
\label{results_all_shadow}
\end{figure}

Another system parameter that can vary randomly in a realistic propagation scenario is the effect of shadowing. The robustness of the proposed learning-based resource allocation scheme is checked by varying the height of the shadowing object when perfect user position information is available. The same evaluation methodology is applied, as done for the case of robustness evaluation of the proposed scheme for varying scatterers' density. Figure \ref{results_all_shadow} shows the performance of the proposed scheme compared to the optimal system performance when the height of shadow object is increased from 1.5 m to 5.0 m. Here, again, we observe that the performance loss does not vary significantly; maximum loss of about 5\% is observed, when the shadowing effect is increased by increasing the height of the shadow object. Since LOS is existent at all times between the users and their corresponding RRHs, therefore, the channel coefficients do not vary significantly with the variation in shadowing effect, which in turn does not affect the transport capacity per user, and hence, the overall sum-goodput of the system. 

\subsection{Effect of Overhead on Throughput of a 5G System}
\label{results_OH}
\begin{figure}
\centering
\includegraphics[width=8.5cm,height=5cm,keepaspectratio]{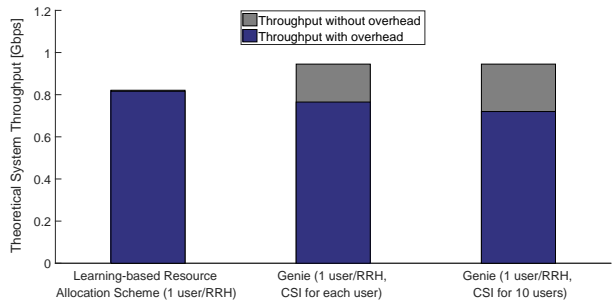}
\caption{Theoretical system throughput for a 5G CRAN system for different schemes in different scenarios}
\label{results_TP_5G}
\end{figure}

After comparing the performance of the proposed learning-based scheme with the traditional CSI-based scheme for resource allocation, we now consider the effect of overhead on the overall system performance for 5G CRAN. Figure \ref{results_TP_5G} shows the theoretical system throughput considering the parameter settings for a TDD-based 5G system. It can be seen that the learning-based resource allocation scheme, considering the simulation scenario in figure \ref{sim_scenario}, does not suffer from the inclusion of the system overhead, where 4 RRHs serve 1 user each, after acquiring their position information. However, the theoretical system throughput for the same scenario using the traditional CSI-based resource allocation scheme is reduced by almost 19\% considering the effect of the CSI acquisition overhead. In a realistic scenario, there are more users lying close to the user served by an RRH, such that the RRH has to acquire CSI for all those users in order to optimally serve the intended user. In this case, the effect of CSI acquisition overhead further increases to about 25\%. The overhead for each of the cases is computed by keeping in mind the assignment of CSI acquisition pilots based on the cyclic-prefix compensation distance, as mentioned in section \ref{OH_model}. Overall, it can be seen that the overhead for CSI acquisition increases with the number of users, thus decreasing the effective system throughput, whereas for position acquisition, the overhead will not impact the effective system throughput since only narrow-band beacons are sufficient for obtaining the position information for the users to be served by a given RRH.

\section{Conclusion}
\label{con}
This paper proposed a novel learning-based resource allocation scheme for 5G CRAN systems, which allocates the system resources based on only the position information of the users present in the system. An overhead model is also presented, for both the position information and CSI acquisition of the users, and its effect on system performance is evaluated. The operation of the proposed scheme based on usage of only the positioning beacons avoids the CSI acquisition overhead, while achieving close to optimal system performance. Overall, less than 15\% loss in system goodput is observed when the proposed scheme is used for resource allocation, compared to the optimal CSI-based resource allocation scheme. However, the proposed scheme has an overhead of only 2.4\% for the presented simulation scenario, compared to an overhead of about 19\% for the CSI-based scheme, and thus, has a better performance in terms of effective system throughput. The proposed scheme is robust to realistic system changes as well, where the maximum performance loss of about 30\% is observed for the case when the reported user's position information has an inaccuracy variance of $1.0$ m. The proposed resource allocation scheme is fairly robust to the changes in the propagation environment; maximum performance loss of 5\% is observed when the system parameters affecting the scattering and shadowing phenomena are different for the training and test datasets used for the machine learning unit of the learning-based resource allocation scheme. The performance loss for inaccurate position information availability can be reduced by using restricted combinations of transmit beam and receive filters (for a given user position) while training the machine learning unit of the proposed scheme, which is included in the related future work. Furthermore, the performance of the proposed scheme can be evaluated when inter-user interference is present in addition to the cross-channel interference, or for different transmit power settings, or when LOS link is not ensured at all times between the RRHs and the users, in the 5G CRAN system.


%
\bibliographystyle{abbrv}
\bibliography{ref_sahar}  
%
\end{document}